\documentclass[aps,nofootinbib,superscriptaddress, showpacs,preprintnumbers,  nofootinbibt,twocolumn]{revtex4-2}
\usepackage{epsfig}
\usepackage{multirow}
\usepackage{eurosym}
\usepackage{dcolumn}
\usepackage{bm}
\usepackage{enumerate}
\usepackage{float}
\usepackage{epstopdf}
\usepackage{amsmath}
\usepackage{bm}
\usepackage{amsfonts}
\usepackage{amssymb}
\usepackage{graphicx}
\usepackage{alphalph,mathtools}
\usepackage{etoolbox}
\usepackage{color}
\usepackage{booktabs}
\usepackage{footnote}
\usepackage{makecell,tabularx}
\usepackage{hyperref}
\hypersetup{colorlinks,citecolor=blue}
\hypersetup{colorlinks=true,linkcolor=red,filecolor=magenta,
    urlcolor=blue}

\def\be{\begin{equation}}
    \def\ee{\end{equation}}
\def\bea{\begin{eqnarray}}
    \def\eea{\end{eqnarray}}

\begin{document} 

\title{\bf Cosmological Tests of $f(R,G,\mathcal{T})$ Dark Energy Model in FRW Universe}

\author{Himanshu Chaudhary}
\email{himanshuch1729@gmail.com} \affiliation{Department of Applied Mathematics, Delhi Technological University, Delhi-110042, India,}
\affiliation{Pacif Institute of Cosmology and Selfology (PICS), Sagara, Sambalpur 768224, Odisha, India}
\affiliation{Department of Mathematics, Shyamlal College, University of Delhi, Delhi-110032, India.}
\author{Amine Bouali}
\email{a1.bouali@ump.ac.ma}
\affiliation{Laboratory of Physics of Matter and Radiation, Mohammed I University, BP 717, Oujda, Morocco,}
\author{Niyaz Uddin Molla}
\email{niyazuddin182@gmail.com}
\author{Ujjal Debnath}
\email{ujjaldebnath@gmail.com} \affiliation{Department of Mathematics, Indian Institute of Engineering Science and Technology, Shibpur, Howrah-711 103, India.}
\author{G.Mustafa}
\email{gmustafa3828@gmail.com} \affiliation{Department of Physics,
Zhejiang Normal University, Jinhua 321004, Peoples Republic of
China,}
\affiliation{New Uzbekistan University, Mustaqillik ave. 54, 100007 Tashkent, Uzbekistan,}

\begin{abstract}
The paper is devoted to study the observational signatures of $f(R,G,\mathcal{T})$ Gravity in FRW Universe. In this research article, we present a new cosmological model formulated within the $f(R,G,\mathcal{T})$ framework. To constrain the model parameters, we employ the Markov Chain Monte Carlo (MCMC) technique, which enables us to explore the parameter space effectively, and used the 36 points of Cosmic Chronometers and 1701 points from Pantheon Plus data. We compare our proposed $f(R,G,\mathcal{T})$ model with the widely accepted $\Lambda$CDM model, considering different cosmological parameters, including Deceleration, Snap, and Jerk. By evaluating these parameters, we gain valuable insights into the dynamics and evolution of the universe within the context of our new model. Moreover, various diagnostic tests have been conducted, such as Statefinder and $Om$ Diagnostic, to further investigate the behavior and consistency of our $f(R,G,\mathcal{T})$ model. These tests offer deeper insights into the properties of our model and its compatibility with observational data. We subject our model to statistical analysis using Information Criteria, which serves as a rigorous quantitative assessment of the model's goodness of fit to the data. This analysis aids in determining the level of agreement between our $f(R,G,\mathcal{T})$ model and the observational data, thus establishing the viability and reliability of our proposed cosmological framework. Our findings highlight the potential of the $f(R,G,\mathcal{T})$ framework in understanding the fundamental aspects of the universe's evolution and dynamics. The comparative analysis with $\Lambda$CDM, as well as the comprehensive diagnostic tests performed, demonstrate the efficacy and validity of our model in explaining the observed cosmological phenomena. These results contribute to the ongoing pursuit of accurate and comprehensive models that can provide a deeper understanding of the nature of our universe.
\end{abstract}

\pacs{}
\maketitle
\tableofcontents

\clearpage
\section{Introduction}
Recent astronomical observations have strongly suggested that our
universe is undergoing an accelerated expansion \cite{R1,P1,S1,S2}
and some unknown matter causes of this acceleration called  Dark
energy (DE) which is basically the negative pressure and positive
energy density form of matter satisfying $\rho+3P<0$ \cite{B1,C1}.
This  DE is one of the most interesting and important discovery of
the mysterious energy in modern cosmology, which was first
investigated in type Ia Supernovae (SNe Ia)\cite{SP1,SP2,SP3} and
afterwards others numerous investigations were done with further
astronomical data \cite{CF1,AGR1,RRC1,SFD1,ZYH1}.It has also
been investigated from the different point of view such as
observational Hubble parameter data,power spectra, cosmic
microwave background radiation(CMBR) and large-scale structure of
the new version of the universe \cite{R2,P2,B2,GH2,E1}. It has
been suggested that the new version of the universe is made of
about $74\% $ DE , $4\% $ ordinary matter and about  $22\% $ dark
matter.The most simplest form of DE  is known as cosmological
constant \cite{TP1,VS1}, also called the $\Lambda$
cold-dark-matter ($\Lambda$CDM) model, which has been successfully
employed to describing the various aspects of the observed
universe. For a long time, researchers on the universe mainly
motivated by theories.However, huge progresses on observational/
experimental explorations of accelerated expansion of the universe
have been witnessed in the last few years. Now a days , the modern
researchers have a great challenging task to know the
nature/properties of the new version of the universe in details.
In the investigation  of the accelerated expansion of the
universe, several cosmological model have been explored mainly in
two different ways. One of them is modified theories of gravity as
a classical  modification of Einstein theory of  gravity and other
approaches is the existence of mysterious energy so-called DE. It
is worth suggesting that the new version of the universe is while
accelerating , the value of EoS parameter (ratio between pressure
and its energy density) is less than $ <-\frac{1}{3}$ and also by
the  observational point of view , the value of the EoS parameter
is  very close to $ -1 $. For this purpose, several works have
been studied by some authors  to describe the DE on several topics
such as quintessence\cite{B3,L1,K1,K2}, quintom \cite{G1,MRS8,A1},
cosmological constant \cite{W1,C2,S3,S4} ,phantom
\cite{S5,S6,P3,K3,S7,K4}  ,tachyon \cite{I6,F1}, modified gravity
\cite{K6,S8,S9,W2,A5,L2,D2,H2}, teleparallel gravity
\cite{S16,P6}, Chaplygin gas models
\cite{A2,A3,N1,K5,A4,P4,P5},holography
\cite{F2,S10,M2,K7,K8,Z2,MB3}, new agegraphics
\cite{K9,S11,S12,S13}, bouncing theory \cite{S14,ARA6,S15} and
braneworld models \cite{S17,DB3}.\\\\
An enormous  relativistic astrophysicist have been proposed the
various modified gravity theories which are obtained via changing
the ricci scalar R in standard Einstein-Hilbert(EH) action. The
standard EH action is changed by the different function of ricci
scalar (i.e. f(R) gravity )\cite{KB1,SP5}or Gauss-Bonnet (GB)
invarient $G$ (i.e.f(G) gravity)\cite{MS2,SN2}. The formulation of
this modifications could be applied as an important role to
investigating the cases of cosmic rapid expansion
\cite{J1,S18,KB2,K10,CD1,KB3,KB4,Y1,Y2,N4,N5,N6,H3,C4}. Without
considering any other dark components ,these modification could
also be described the early-time inflation as well as late time
acceleration  and these cosmological model could also be
consistent with the solar system constraint \cite{DF6}. Modified
gravity theories refer to the alternative theories of gravity that
differ from Einstein's general theory of relativity (GR) by
modifying the fundamental gravitational action. In particular,
$f(R)$ gravity \cite{N2}, where $R$ is the Ricci scalar, is a popular
modified gravity theory that has received significant attention in
recent years due to its ability to explain the current cosmic
acceleration without invoking dark energy. However, $f(R)$ gravity
alone cannot explain all of the observed phenomena in astrophysics
and cosmology. A class of generalized $f(R)$ modified theories of
gravity derived by Bertolami et. al \cite{B5} by considering
an explicit coupling between of an arbitrary function of the Ricci
scalar curvature $R$ and the matter Lagrangian density
$\mathcal{L}_m$. In the description of early-time and late-time
accelerated universe, {\color{red} Nojiri} et al. \cite{N1,B6} investigated the
non-minimal coupling of $f(R)$ and $f(G)$ gravity of theories with the
Lagrangian density of matter $\mathcal{L}_m$ and shown that the
unified description of the inflationary era with the present
cosmic accelerated expansion. As an alternative to DE, to study
the various cosmic issues, an systemetic method has discussed by
the $f(G)$ gravity \cite{N8} and it is extremely useful to study the
restricted period of the future discontinuties and the pace of
cosmos over the long period time \cite{N7,B8}. Harko et al
\cite{H4} introduced another important extension of general theory
relativity such as $f(R,T)$ and f(R,$T^{\phi}$) where the
gravitational Lagrangian  is the function of ricci scalar R and
trace of stress energy tensor T where as $T^{\phi}$ for the scalar
field stress energy tensor. Several important implication in
$f(R,T)$ gravity theory have been extremely  discussed in some
Litaratures \cite{J3,H5,S21,A7,M11,C5}.Later, Researchers have
introduced the another type of modified gravity such as $f(R,G)$
gravity \cite{B7,N3}. Several cosmological implication including the
energy conditions,future finite-time singularities have been
discussed in several literatures
\cite{DF1,DF2,DF3,DF4,CD3,M10,A6,E4}. In \cite{S19}, Sharif and
Ikram proposed another kind of extension of modified theory of
gravity like $f(G,T)$ gravity theory. They reconstructed the $f(G,T)$
gravity theory through the power-law, de-sitter expansion of the
universe. They also studied the stability of some reconstructed
cosmological model with liner perturbations in $f(G,T)$  Shamir and
Ahmed \cite{S20} investigated the $f(G,T)$ gravity using the Noether
symmetry of some cosmological viable. \\\\
In the ref.\cite{UD}, the author proposed the modification of
above modified gravities, named as $f(R,G,\mathcal{T})$ gravity.
The $f(R,G,\mathcal{T})$ gravity is a more general modification of
gravity, where $G$ and $\mathcal{T}$ are the Gauss-Bonnet
invariant and the trace of the energy-momentum tensor,
respectively. It has been found the forms of the function
$f(R,G,\mathcal{T})$ by the three types of standard expansion
models such as de Sitter, power-law and future singularity models.
This theory is considered more promising because it can address
not only the cosmic acceleration but also several other problems
in the field of astrophysics and cosmology. One of the most
significant implications of $f(R,G,\mathcal{T})$ gravity is its
ability to explain the current cosmic acceleration without
introducing dark energy. Recent observational studies have
confirmed that the expansion rate of the universe is increasing,
but the cause of this acceleration is not yet fully understood. In
standard cosmology, dark energy is invoked to explain this
phenomenon. However, $f(R,G,\mathcal{T})$ gravity provides a
viable alternative explanation without introducing any new unknown
physical entity. Moreover, $f(R,G,\mathcal{T})$ gravity can also
address some of the shortcomings of the $f(R)$ gravity. For
example, $f(R)$ gravity is known to produce some inconsistencies
when tested against observations of gravitational waves. However,
$f(R,G,\mathcal{T})$ gravity is free from these inconsistencies,
making it a more attractive theory. In addition to explaining the
cosmic acceleration, $f(R,G,\mathcal{T})$ gravity can also address
the issue of dark matter and the formation of large-scale
structures in the universe. Recent studies have shown that
$f(R,G,\mathcal{T})$ gravity can explain the observed rotation
curves of galaxies without the need for dark matter. Furthermore,
it can also reproduce the observed cosmic microwave background
radiation and large-scale structure formation. While
$f(R,G,\mathcal{T})$ gravity is still a relatively new and
untested theory, it has shown great potential in addressing some
of the most significant problems in the field of astrophysics and
cosmology. Its ability to unify modified gravity theories and
address several issues makes it an important area of research.
Subsequently, some authors \cite{I1,I2,I3,I4} have studied the
wormhole and compact star models in the framework of
$f(R,G,\mathcal{T})$ gravity.\\\\
In the present work, we consider the newly proposed
$f(R,G,\mathcal{T})$ gravity model in FRW universe. The
$f(R,G,\mathcal{T})$ gravity is strongly motivated by its
potential to explain the observed accelerated expansion of the
universe without introducing dark energy, offering a more elegant
and physically intuitive explanation for this cosmic phenomenon.
By unifying various modified gravity theories and satisfying solar
system constraints, it provides a comprehensive framework for
gravity at different scales. Moreover, its ability to address dark
matter, reproduce observed galactic rotation curves, and explain
the formation of large-scale structures in the universe
underscores its versatility and relevance in solving multiple
astrophysical mysteries. Additionally, $f(R,G,\mathcal{T})$
gravity offers insights into gravitational wave behavior without
the inconsistencies encountered by other modified gravity
theories, making it a promising avenue for advancing our
understanding of gravity in extreme environments. Overall, this
theory represents a compelling and promising approach to
addressing fundamental questions in cosmology and astrophysics
while simplifying our cosmic model.\\\\
The paper is organized as follows: In section
\ref{sec1}, we assume the universe is filled with radiation and
dark matter in the framework of $f(R,G,\mathcal{T})$ gravity. We
are not taking any external dark energy where the effect of this
modification of gravity can be treated as alternative to dark
energy. Then we assume a power-law form of the function
$f(R,G,\mathcal{T})$ and then form a differential equation of the
Hubble parameter $H(z)$. In section \ref{sec2}. we constrain the
parameters of the $f(R,G,\mathcal{T})$ gravity model by MCMC
method and then obtain the viability of the model. In Section
\ref{sec3} compares the model's predictions with observational
data. Section \ref{sec4} gives a detailed description about the
kinematic cosmographic parameters such as the deceleration, jerk
and snap parameters. Section \ref{sec5} and \ref{sec6} discuss
about the statefinder and Om diagnostics and present the evolution
history of dark energy on $s-r$ and $q-r$ planes. Section
\ref{sec7} discusses about the information criteria. Section
\ref{sec8} and Section \ref{sec10} discuss the results and conclusions  respectively.

\section{Basic Equations in $f(R,G,\mathcal{T})$ Gravity}\label{sec1}
In this section, we review the $f(R,G,\mathcal{T})$ gravity theory in details.
The HE action for $f(R,G,\mathcal{T})$ gravity theory is defined in
the form \cite{UD}
\begin{equation}\label{1}
S=\frac{1}{2}~\int f(R,G,\mathcal{T})\sqrt{-g}~d^{4}x +\int L_{m}
\sqrt{-g}~d^{4}x
\end{equation}
where $f(R,G,\mathcal{T})$  denotes the arbitrary function of $R$,  $G$ and
$\mathcal{T}$  where  $L_{m}$ denotes the matter Lagrangian, $g=|g_{\mu\nu}|$ ($g_{\mu\nu}$ is the
metric tensor) (choosing $8\pi G_{N}=c=1$, $G_{N}$ is the Newtonian constant). The Ricci scalar $R$ , Gauss-Bonnet
invariant $G$  and the trace of  stress energy tensor $\mathcal{T}$  are defined as follows:
\begin{equation}\label{5}
R=g^{\mu\nu}R_{\mu\nu},~G=R^{2}-4R_{\mu\nu}R^{\mu\nu}+R_{\mu\nu\xi\eta}R^{\mu\nu\xi\eta},~\mathcal{T}=g^{\mu\nu}g T_{\mu\nu}
\end{equation}
The  desired field equations of $f(R,G,\mathcal{T})$ gravity  theory are obtained by the variation of action (\ref{1}) as follows
$$
(R_{\mu\nu}+g_{\mu\nu}\nabla^{2}-\nabla_{\mu}\nabla_{\nu})f_{R}-\frac{1}{2}~fg_{\mu\nu}
$$
$$+(2RR_{\mu\nu}-4R^{\xi}_{\mu}R_{\xi\nu}-4R_{\mu\xi\nu\eta}R^{\xi\eta}
+2R_{\mu}^{\xi\eta\lambda}R_{\nu\xi\eta\lambda})f_{G}
$$
$$
+(2Rg_{\mu\nu}\nabla^{2}-2R\nabla_{\mu}\nabla_{\nu}-4g_{\mu\nu}R^{\xi\eta}\nabla_{\xi}\nabla_{\eta}
$$

$$
-4R_{\mu\nu}\nabla^{2}+4R_{\mu}^{\xi}\nabla_{\nu}\nabla_{\xi}+4R_{\nu}^{\xi}\nabla_{\mu}\nabla_{\xi}
+4R_{\mu\xi\nu\eta}\nabla^{\xi}\nabla^{\eta})f_{G}
$$
\begin{equation}\label{13}
=T_{\mu\nu}-(T_{\mu\nu}+\Theta_{\mu\nu})f_{\mathcal{T}}
\end{equation}
where $f_{R}=\frac{\partial f}{\partial R}$, $f_{G}=\frac{\partial
f}{\partial G}$, $f_{\mathcal{T}}=\frac{\partial f}{\partial
\mathcal{T}}$, $\nabla^{2}=\nabla_{\mu}\nabla^{\mu}$ is the
D'Alembert operator, $T_{\mu\nu}$ is the energy-momentum tensor
and $\mathcal{T}$ is its trace.
The energy momentum tensor for
an ideal fluid is represented as
\begin{equation}
T_{\mu\nu}=(\rho+p)u_{\mu}u_{\nu}+pg_{\mu\nu}
\end{equation}
where  the symbol $\rho$ and $p$ denote  the energy density and pressure
of  the ideal fluid respectively . The four velocity  of the fluid's $u_{\mu}$ satisfies
$u_{\mu}u^{\mu}=-1$ and $u^{\mu}\nabla _{\nu}u_{\mu}=0$. Also
$\Theta_{\mu\nu}=-2T_{\mu\nu}+pg_{\mu\nu}$.
The line element of the
flat Friedmann-Robertson-Walker (FRW) model of the universe
assumed as
\begin{equation}\label{4b}
ds^{2}=-dt^{2}+a^{2}(t)\left[dr^{2}+r^{2}(d\theta^{2}+sin^{2}\theta
d\phi^{2})\right]
\end{equation}
where $a(t)$ is the scale factor. From equation (\ref{5}), we can
obtain
\begin{equation}
R=6(\dot{H}+2H^{2}),~~G=24H^{2}(\dot{H}+H^{2})
\end{equation}
Here  $H=\dot{a}/a$ denotes the Hubble parameter where $dot$ denotes the
derivative w.r.t. cosmic time $t$.Using the above metric (\ref{4b}) , we get
the trace of $T_{\mu\nu}$ is $\mathcal{T}=3p-\rho$. Now, the basic conservation
equation $\nabla^{\mu}T_{\mu\nu}=0$ for an ideal fluid gives
\begin{equation}\label{22}
\dot{\rho}+3H(\rho+p)=0
\end{equation}
with the help of the equation (\ref{13}), the field equations for
$f(R,G,\mathcal{T})$ gravity are obtained as follows


\begin{align}\label{eq8}
3H^{2}=\frac{1}{f_{R}} \Bigg[ & \rho+(\rho+p)f_{\mathcal{T}}+\frac{1}{2}(Rf_{R}-f) \nonumber \\
& -3H\dot{f}_{R}+12H^{2}(\dot{H}+H^{2})f_{G}-12H^{3}\dot{f}_{G} \Bigg]
\end{align}

and

\begin{align}\label{25}
(2\dot{H}+3H^{2}) = -\frac{1}{f_{R}} \Bigg[ & p-\frac{1}{2}(Rf_{R}-f)+2H\dot{f}{R}+\ddot{f}{R} \nonumber \\
&-12H^{2}+(\dot{H}+H^{2})f_{G} \\
&+8H(\dot{H}+H^{2})\dot{f}{G}+4H^{2}\ddot{f}{G} \Bigg]
\end{align}

In the standard
Einstein's field equations, the above two field equations can be written as
\begin{equation}
3H^{2}=\rho_{eff}~~and~~ (2\dot{H}+3H^{2})=-p_{eff}
\end{equation}
where


\begin{align}
\rho_{eff}=\frac{1}{f_{R}} \Bigg[ & \rho+(\rho+p)f_{\mathcal{T}}+\frac{1}{2}(Rf_{R}-f) \nonumber \\
& -3H\dot{f}_{R}+12H^{2}(\dot{H}+H^{2})f_{G}-12H^{3}\dot{f}_{G} \Bigg]
\end{align}

and

\begin{align}
p_{eff} = \frac{1}{f_{R}} \Bigg[ & p-\frac{1}{2}(Rf_{R}-f)+2H\dot{f}{R}+\ddot{f}{R} \nonumber -12H^{2}\\
& (\dot{H}+H^{2})f_{G} + 8H(\dot{H}+H^{2})\dot{f}{G}+4H^{2}\ddot{f}{G} \Bigg]
\end{align}

We assume that the fluid components of the universe are composed of radiation, pressureless dark matter, and dark energy where dark energy can be produced by modified gravity. So
$\rho=\rho_r+\rho_m$ and $p=p_r+p_m$ with $p_m=0$. The energy
density for radiation and dark matter are respectively
$\rho_{r}=3H_{0}^{2}\Omega_{r0}(1+z)^{4}$ and
$\rho_{m}=3H_{0}^{2}\Omega_{m0}(1+z)^{3}$, where
$\Omega_{r0}=\frac{\rho_{r0}}{3H_{0}^{2}}$ and
$\Omega_{m0}=\frac{\rho_{m0}}{3H_{0}^{2}}$ are the dimensionless
density parameters, $H_{0}$ is the present value of the Hubble
parameter and $z$ is the redshift.\\

 Now we assume a power-law form of $f(R,G,\mathcal{T})$ as
in the form
\begin{equation}
f(R,G,\mathcal{T})=\alpha_{1}R^{\beta_{1}}+\alpha_{2}G^{\beta_{2}}+\alpha_{3}(-\mathcal{T})^{\beta_{3}}
\end{equation}
where $\alpha_i$ and $\beta_i$ ($i=1,2,3$) are constants. So from equation (\ref{eq8}), we obtain the differential equation of the Hubble
parameter $H(z)$:

\begin{widetext}
\begin{eqnarray}
&&24^{\beta _2} \alpha _2 (1-\beta_2) \left(H^3 \left(H-(1+z)
H'\right)\right)^{\beta _2} \nonumber
\\
&&  +6^{\beta _1} \alpha _1 \beta _1 H^2 \left(H \left(2 H-(1+z)
H'\right)\right)^{\beta_1 -1}+6^{\beta _1} \alpha _1 (1-\beta_1)
\left(H \left(2 H-(1+z) H'\right)\right)^{\beta _1} \nonumber
\\
&& +6^{\beta _1} (1+z)  \alpha _1 \beta _1 \left(\beta _1
-1\right) H^3 \left(H \left(2 H-(1+z) H'\right)\right)^{\beta_1
-2} \left((1+z) H'^2-H \left(3 H'-(1+z) H''\right)\right)
\nonumber
\\
&&   +24^{\beta _2} (1+z) \alpha _2 \beta _2 \left(\beta _2
-1\right) H^{3\beta_2} \left(H-(1+z) H'\right)^{\beta _2-2}
\left(3 (1+z) H'^2+H \left(-3 H'+(1+z) H''\right)\right) \nonumber
\\
&&=-2\alpha_3
H_{0}^{2}\left(3H_{0}^{2}\Omega_{m0}(1+z)^{3}\right)^{\beta_3 -1}
\left[3(\beta_3 -0.5)\Omega_{m0}(1+z)^{3}+
4\beta_{3}\Omega_{r0}(1+z)^{4} \right] \nonumber
\\
&& +6H_{0}^{2}\left[ \Omega_{m0}(1+z)^{3}+\Omega_{r0}(1+z)^{4}
\right]
\end{eqnarray}
\end{widetext}

where $H'=dH/dz$ and $H''=d^{2}H/dz^2$. If $\beta_1$ and $\beta_2$ close to zero, $f(R,G,\mathcal{T})$ gravity is converted to $f(R)$ gravity. If $\beta_3$ goes to zero, there will be no effect of $\mathcal{T}$

\section{Cosmological tests of the $f(R,G,\mathcal{T})$ dark energy model}\label{sec2}
In this section, we conduct a comprehensive comparison between the predictions of the $f(R, G,\mathcal{T})$ dark energy model and observational data to obtain constraints on the model's free parameters. Our analysis employs two observational datasets, namely the Cosmic chronometers data and the Pantheon $+$ dataset, which consists of 36 and 1701 data points, respectively. The nine free parameters of the $f(R, G,\mathcal{T})$ dark energy model, namely $\left(H_0,\Omega_{\mathrm{m0}},\Omega_{\mathrm{r0}},\beta_{1},\beta_{2},\beta_{3}, \alpha_{1}, \alpha_{2},\alpha_{3}\right)$, along with the present-day value of the Hubble function $H_0$, are constrained using the standard Bayesian technique, likelihood function approach, and Markov Chain Monte Carlo (MCMC) method. Once the best-fit values of the model parameters are obtained, we examine the model's cosmographic behavior by analyzing the evolution of the deceleration, jerk, and snap parameters, and compare the model predictions with those of the standard $\Lambda$CDM cosmological model.\\\

\subsection{Methodology}
Markov Chain Monte Carlo (MCMC) is a widely used statistical technique in cosmology for exploring the parameter space of complex models and generating probability distributions for cosmological parameters \cite{cosmo_mcmc}. MCMC is particularly useful when there is a large parameter space and the likelihood function is non-Gaussian or non-linear \cite{cosmo_mcmc_brooks_gelman}. The basic idea behind MCMC is to create a Markov chain that samples the parameter space of a model according to a probability distribution. The chain consists of a sequence of parameter values, where each value is generated from the previous value using a set of transition rules that depend on a proposal distribution \cite{cosmo_mcmc_handbook}. The proposal distribution suggests a new parameter value that may or may not be accepted based on its posterior probability given the data and the prior probability distribution \cite{cosmo_mcmc_handbook}. To ensure that the Markov chain converges to the true posterior distribution, several methods are used, including adjusting the proposal distribution to optimize the acceptance rate, tuning the length of the chain to obtain independent samples, and assessing the convergence of the chain using diagnostic tests \cite{cosmo_mcmc_brooks_gelman,cosmo_mcmc_handbook}. In our research, we harnessed the power of the PolyChord algorithm to facilitate our MCMC analysis. PolyChord is an advanced tool that seamlessly integrates with MCMC techniques and aids in efficiently exploring the parameter space. It automates many aspects of the sampling process, streamlining the calculation of evidence (marginal likelihood) and posterior distributions while minimizing the need for manual intervention \cite{pypolychord1,pypolychord2,pypolychord3,pypolychord4}. For the visualization and interpretation of our results, we employed ChainConsumer, a versatile tool for plotting and analyzing MCMC chains. ChainConsumer \cite{chainconumer} enhances the presentation of parameter estimates, posterior distributions, and credible intervals, providing a comprehensive view of the results. Once the chain has converged, the posterior distribution for the parameters can be estimated by computing the frequency of the parameter values in the chain \cite{cosmo_mcmc_handbook}. The posterior distribution can then be used to estimate the best-fit values and uncertainties for the cosmological parameters and to make predictions for observables such as the cosmic microwave background radiation and the large-scale structure of the universe \cite{cosmo_mcmc,cosmo_mcmc_brooks_gelman}. MCMC is a powerful tool for analyzing cosmological data and has been used in a wide range of cosmological studies, including measurements of the cosmic microwave background, large-scale structure, and dark energy \cite{cosmo_mcmc,cosmo_mcmc_brooks_gelman}. The technique is computationally intensive, but with the increasing power of modern computing, it has become an essential tool for cosmologists in understanding the nature and evolution of the universe \cite{cosmo_mcmc,cosmo_mcmc_brooks_gelman}.

\subsection{MCMC Setup and Analysis}
In our research, we used the PolyChord algorithm to explore the complex parameter space of our cosmological model through Markov Chain Monte Carlo (MCMC) analysis. By setting "nlive" to 100, we balanced computational efficiency and result accuracy, enabling a thorough exploration of parameters. We assessed convergence, a crucial MCMC step, by tracking indicators like logZ changes and live point behavior. These indicators stabilized, with logZ. We employed a custom ``uniprior" function for uniform parameter priors and benefited from PolyChord's nested sampling strategy, reducing the need for explicit proposal matrices. For plotting, we leveraged Chain Consumer, streamlining posterior distribution analysis and enhancing the reliability of our cosmological predictions.
\subsection{Data description}\label{DD}

\subsubsection{Cosmic chronometers Dataset}
Cosmic chronometers are a class of astronomical objects that provide valuable data for measuring the expansion history of the universe. These objects are typically old, quiescent galaxies that have stopped forming stars and are identified by their spectra and colors \cite{CC1}. The data used in cosmic chronometry is based on measurements of the ages of these galaxies at different redshifts. Elliptical galaxies are the most commonly used cosmic chronometers since they have relatively simple stellar populations and are thought to have formed early in the history of the universe \cite{CC3}. The ages of stars in these galaxies can be determined from their spectra, which provide information on the chemical composition and internal processes that have occurred within them \cite{CC4}. Other potential cosmic chronometers include white dwarfs, globular clusters, and the oldest stars in the Milky Way. However, these objects are more challenging to observe and analyze than elliptical galaxies \cite{CC5}, and their use as cosmic chronometers is still an area of ongoing research. To collect cosmic chronometry data, astronomers use large telescopes and spectrographs to measure the spectra and colors of these galaxies with high precision \cite{CC2}. Sophisticated statistical techniques are then employed to determine the ages of the galaxies and the expansion history of the universe \cite{CC6}. Ccosmic chronometry provides an essential tool for studying the universe's expansion history and testing cosmological models \cite{CC1}. The data obtained from cosmic chronometers is crucial to our understanding of the universe and the fundamental laws of physics that govern it.\\\\

In our analysis, we incorporated Hubble expansion rate data to obtain tighter constraints on our dark energy (DE) models. The CC dataset can be derived through different methods. One approach is based on the clustering of galaxies and quasars, where the Baryon Acoustic Oscillations (BAO) in the radial direction are measured \cite{CC7,CC8}. Another method is the differential age method, which expresses the Hubble parameter as

\begin{equation}
H(z) = -\frac{1}{1+z} \frac{dz}{dt},
\end{equation}

where $\frac{dz}{dt}$ can be inferred from $\frac{\Delta z}{\Delta t}$. Here, $\Delta z$ and $\Delta t$ represent the redshift difference and the age difference between two passively evolving galaxies, respectively \cite{CC9,CC10}. For our analysis, we utilized a compilation of 36 data points of the CC dataset, with each data point accompanied by its corresponding reference. While the CC data points are considered uncorrelated, we can define the $\chi_{H(z)}^2$ function as

\begin{equation}
\chi_{CC}^2 = \sum_{i=1}^{36} \left[\frac{H_{\text{obs},i} - H(z_i)}{\sigma_{H,i}}\right]^2,
\end{equation}

where $H_{\text{obs}, i}$ represents the observed value of the Hubble parameter for each redshift $z_i$ (references), and $H(z)$ denotes the theoretical prediction of the Hubble parameter. By evaluating the $\chi_{CC}^2$ function, we can quantify the level of agreement between the observed Hubble parameter values and the predictions of our DE models. This analysis allows us to assess the fit quality and obtain valuable constraints on the parameters of the models.

\subsubsection{Pantheon + Dataset}
The updated Pantheon $+$ dataset represents a significant advancement in our understanding of the universe's expansion history. This compilation incorporates a comprehensive collection of 1701 data points, providing a wealth of information for cosmological investigations. These data points cover a wide range of redshifts, from $0.001 < z < 2.3$ allowing researchers to probe the expansion dynamics over a significant cosmic timeline. The Pantheon $+$  dataset builds upon previous SNIa compilations and includes the latest observations of Type Ia supernovae. These supernovae have been instrumental in unveiling the accelerating expansion of the universe. As highly luminous astrophysical objects, SNIa serves as valuable standard candles for measuring relative distances based on their apparent and absolute magnitudes.\\\\

Type Ia supernovae (SNIa) have played a significant role in our understanding of the accelerating expansion of the universe. These astrophysical objects have proven to be valuable tools for studying the nature of the component responsible for this cosmic acceleration. Several compilations of SNIa data have been released in recent years, such as Union \cite{SupernovaCosmologyProject:2008ojh}, Union2 \cite{Amanullah:2010vv}, Union2.1 \cite{SupernovaCosmologyProject:2011ycw}, Joint Light-curve Analysis (JLA) \cite{SDSS:2014iwm}, Pantheon \cite{Pan-STARRS1:2017jku}, and the more recent Pantheon+ \cite{Scolnic:2021amr}. The Pantheon+ dataset, which contains 1701 SNIa spanning a redshift range of $0.001 < z < 2.3$, provides a valuable resource for cosmological investigations. SNIa are exceptionally luminous objects and are often considered as standard candles for measuring relative distances in the universe using the distance modulus. The chi-square values associated with the Pantheon+ dataset are calculated as follows:

\begin{equation}\label{eq_chi2}
\chi^2_{\text{Pantheon+}} = \vec{D}^T \cdot \mathbf{C}^{-1}_{\text{Pantheon+}} \cdot \vec{D},
\end{equation}

where $\mathbf{C}_{\text{Pantheon+}}$ represents the covariance matrix provided with the Pantheon+ data, encompassing both statistical and systematic uncertainties. In Equation (\ref{eq_chi2}), $\vec{D} = m_{B i} - M - \mu_{\text{model}}$, with $m_{B i}$ and $M$ denoting the apparent and absolute magnitudes, respectively. The term $\mu_{\text{model}}$ represents the predicted distance modulus based on a chosen cosmological model, given by

\begin{equation}
\mu_{\text{model}}(z_{i}) = 5\log_{10}\left(\frac{D_L(z_{i})}{\frac{H_0}{c}\text{Mpc}}\right) + 25,
\end{equation}

where $H_0$ is the present value of the Hubble rate, and $D_L(z)$ denotes the luminosity distance. For a flat, homogeneous, and isotropic FLRW universe, $D_L(z)$ is given by

\begin{equation}
D_L(z) = (1+z)H_0 \int_{0}^{z}\frac{dz^{\prime}}{H(z^{\prime})}.
\end{equation}

Unlike the Pantheon dataset, the degeneracy between the absolute magnitude $M$ and $H_0$ is broken in Pantheon+. This is achieved by rewriting the vector $\vec{D}$ in Equation (\ref{eq_chi2}) in terms of the distance moduli of SNIa in Cepheid hosts. This allows for an independent constraint on $M$, resulting in the following expression:

\begin{equation}
\vec{D}_{i}^{\prime}= \begin{cases}m_{B i}-M-\mu_{i}^{\text {Ceph }} & i \in \text { Cepheid hosts } \\ m_{B i}-M-\mu_{\text {model }}\left(z_{i}\right) & \text { otherwise },\end{cases}
\end{equation}

where $\mu_{i}^{\text{Ceph}}$ represents the distance modulus corresponding to the Cepheid host of the $i^{\text{th}}$ SNIa, measured independently using Cepheid calibrators. Thus, Equation (\ref{eq_chi2}) can be rewritten as:

\begin{equation}\label{eq_chi2_rew}
\chi^2_{\text{SN}} = \vec{D}^{\prime T} \cdot \mathbf{C}^{-1}_{\text{Pantheon+}} \cdot \vec{D}^{\prime}.
\end{equation}
\\\\

In order to combine the SNIa data with other cosmological probes, the total chi-square value, $\chi^2_{\text{tot}}$, is obtained by adding the contributions from the cosmic chronometers (CC) and the other cosmological datasets (Pantheon $+$).

\begin{equation}
\chi^{2}_{Tot}=\chi_{CC}^{2}+{\chi}_{{Pantheon+}}^2.
\end{equation}

The contour plots for the combined result of CC + SNIa are shown in the following Fig:- \ref{MCMC} and the best-fit values are tabulated in Table \ref{table1}.

\begin{figure*}[htbp]
\centering
\includegraphics[scale=0.52]{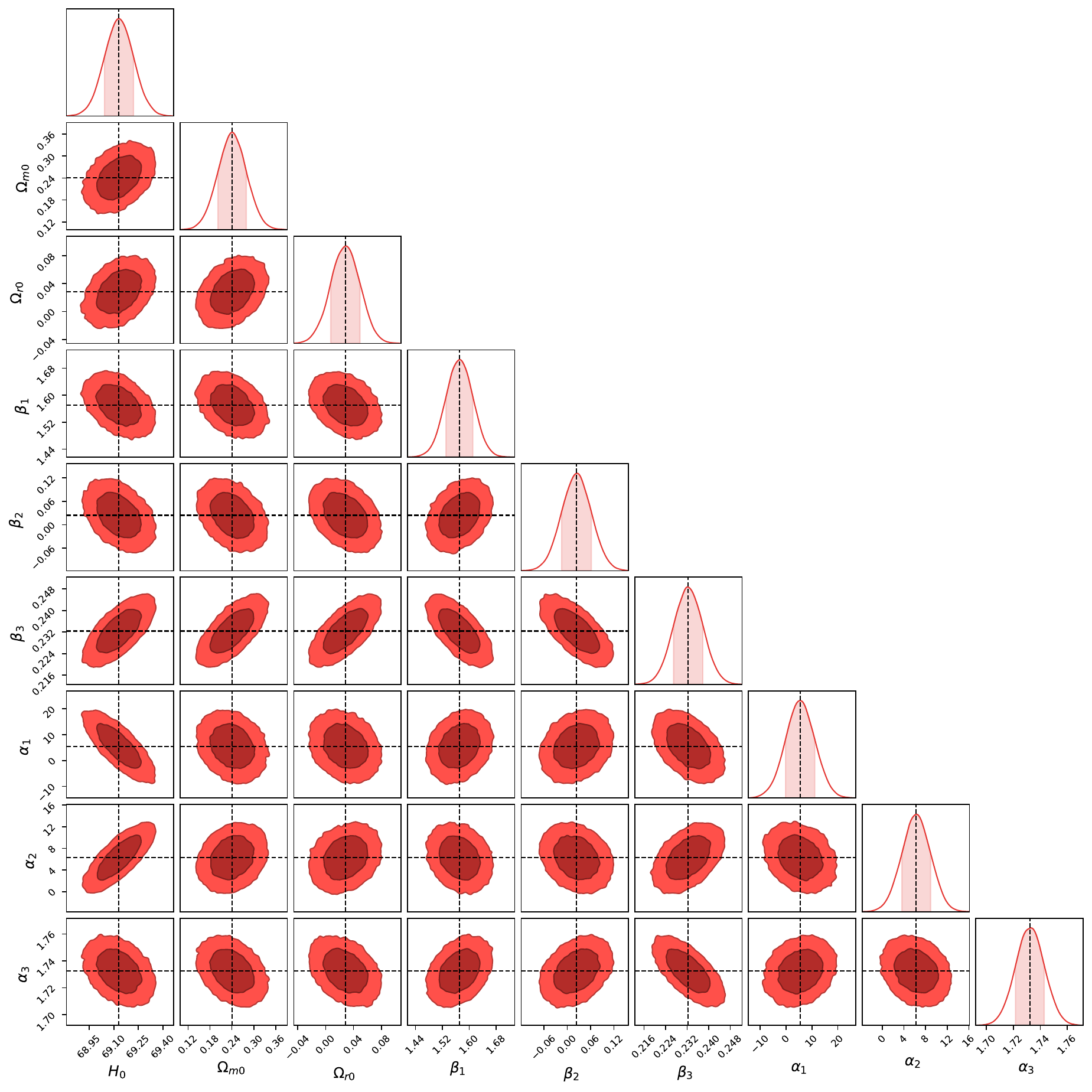}
\caption{The contour plot of free parameter space $(H_0,\Omega_{\mathrm{m0}},\Omega_{\mathrm{r0}},\beta_{1},\beta_{2},\beta_{3}, \alpha_{1}, \alpha_{2},\alpha_{3})$ for our model I with $1-\sigma$ and $2-\sigma$ errors obtained from the  datasets.}\label{MCMC}
\end{figure*}

\begin{table}[H]
\begin{center}
\begin{tabular}{|c|c|c|c|}
\hline
\multicolumn{4}{|c|}{MCMC Results of $f(R,G,\mathcal{T})$ Model} \\
\hline
Model & Parameters & Prior & Best fit Value \\[1ex]
\hline
$  \Lambda$CDM Model  & $H_0$ & [50.,100.] &$69.854848_{-1.259100}^{+1.259100}$ \\[1ex]
\hline
Model & $H_0$& [50.,100.] & $69.131041^{+0.139490}_{-0.139490}$  \\[1ex]
 &$\Omega_{\mathrm{m0}}$  & [0.,1.] & $0.240739 ^{+0.018284}_{-0.018284}$\\[1ex]
 &$\Omega_{\mathrm{r0}}$ & [0.,1] &$0.028080^{+0.009746}_{-0.009746}$ \\[1ex]
& $\beta_{1}$& [1.,2.] &$1.570660^{+0.018798}_{-0.018798}$ \\[1ex]
&$\beta _{2}$& [0.,0.1] &$0.024082_{-0.017751}^{+0.017751}$ \\[1ex]
&$\beta _{3}$& [0.,1]&$0.232398_{-0.004917}^{+0.004917}$
\\[1ex]
&$\alpha_{1}$& [2.,8.] &$5.496010^{+2.820425}_{-2.820425}$ \\[1ex]
& $\alpha_{2}$& [4.,8] &$6.250227^{+1.312755}_{-1.312755}$ \\[1ex]
&$\alpha_{3}$ & [1.,2.] &$1.732398_{-0.004917}^{+0.004917}$ \\[1ex]
\hline
\end{tabular}
\caption{Summary of the MCMC results using
dataset.}
\label{table1}
\end{center}
\end{table}

\section{Observational, and theoretical comparisons of the Hubble functions}\label{sec3}

\subsection{Comparison with the Cosmic chronometers dataSet.}
The Hubble function is a crucial parameter in cosmology, as it relates the universe's expansion rate to its present age and the distribution of matter and energy within it. The $\Lambda$CDM model is the most widely accepted cosmological model, describing the universe as composed of dark matter, dark energy, and ordinary matter. In this study, we perform curve fitting of the Hubble function using the $\Lambda$CDM model, $f(R, G,\mathcal{T})$ dark energy model, and 36 points of Cosmic Chronometers (CC) by using the best-fit values of both the model parameters obtained by minimizing the $\chi^{2}$ function.\\\\

\begin{figure}[!htb]
\centering
\includegraphics[scale=0.42]{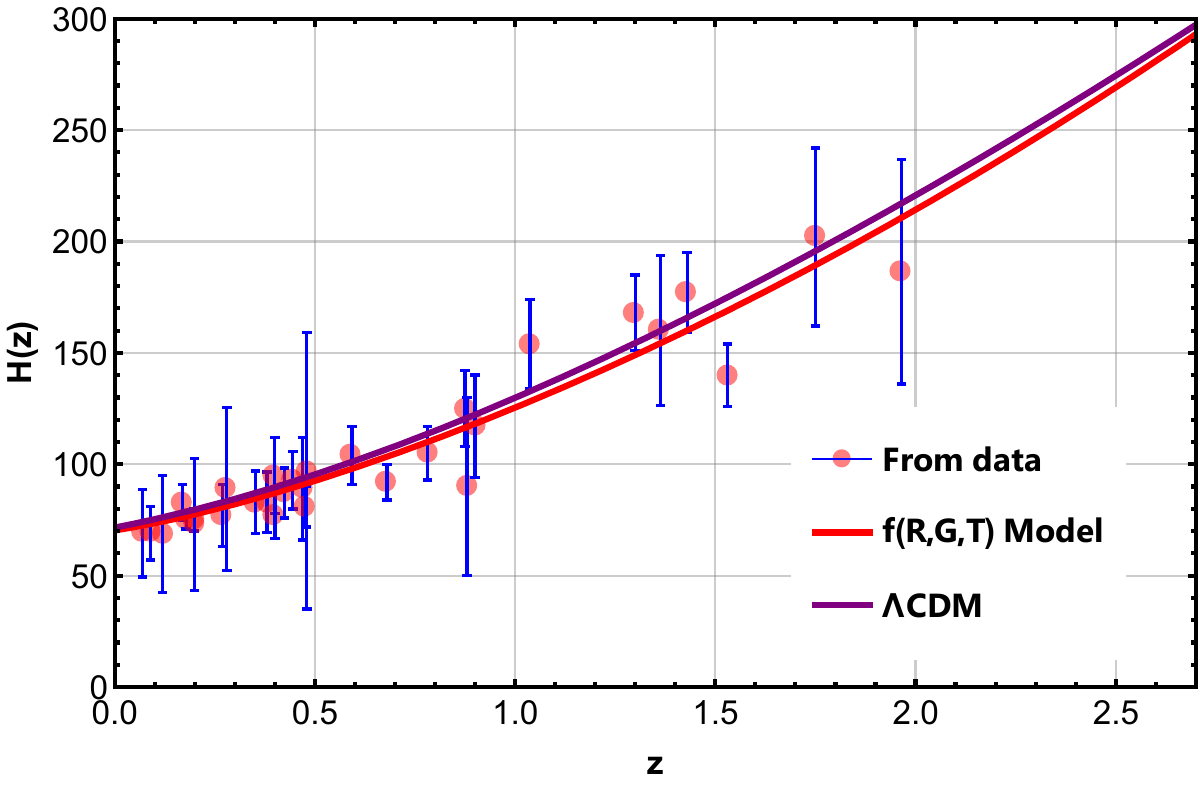}
\caption{The figure shows the theoretical curve of the Hubble function $ H(z) $ corresponding to the studied (red curve) and $ \Lambda$CDM (purple curve) models against the 36 measurements of CC measurements shown in magenta dots with their corresponding error bars in the blue line. For $\Lambda$CDM the parameters were fixed at $\Omega_m=0.301\pm0.012$ and $\Omega_\Lambda=0.699\pm0.012$.}
\label{h(z)}
\end{figure}

Our results demonstrate that the $\Lambda$CDM model provides an excellent fit to the CC data, and $f(R, G,\mathcal{T})$ dark energy model. The comparison findings are shown in Figure \ref{h(z)}.

\subsection{Comparison with the Relative difference between $f(R, G,\mathcal{T})$ dark energy model and $\Lambda$CDM}
\begin{figure}[!htb]
\centering
\includegraphics[scale=0.42]{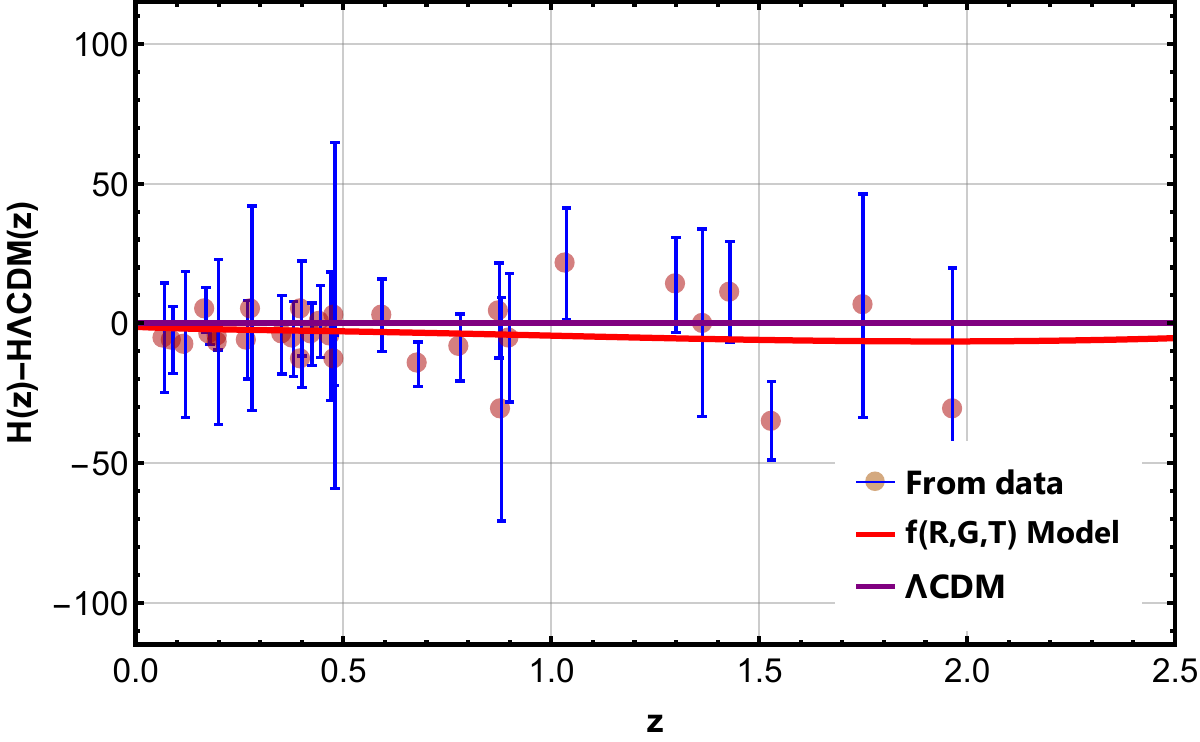}
\caption{The variation of the difference between $f(R, G, T)$ Model corresponding to the studied (red curve)  and $ \Lambda$CDM (purple curve) as a function of the redshift $z$ against the CC measurements in magenta dots with their corresponding error bars in the blue line. For $\Lambda$CDM the parameters were fixed at $\Omega_m=0.301\pm0.012$ and $\Omega_\Lambda=0.699\pm0.012$.}
\label{h(z)diff}
\end{figure}

In this subsection, we aim to study the difference between the $f(R, G,\mathcal{T})$ dark energy model and the $\Lambda$CDM model as a function of redshift, by comparing their predicted values for the Hubble parameter against Cosmic Chronometers (CC) measurements. We then analyze how this difference varies with redshift.\\\\
Our findings indicate that there is a negligible difference between the $f(R, G,\mathcal{T})$ dark energy model and the $\Lambda$CDM model at low redshifts but slightly increases at higher redshifts, However, we acknowledge that the uncertainties in the CC measurements also increase with redshift, so caution must be taken when interpreting these results. Overall, our investigation provides significant insights into the limitations of the $\Lambda$CDM model and the potential of $f(R, G,\mathcal{T})$ dark energy model to better explain the behavior of the universe at higher redshifts. The comparison findings are shown in Figure \ref{h(z)diff}.

\subsection{Comparison with the Pantheon $+$ dataSet}
The distance modulus function has been fitted with the $\Lambda$CDM model and $f(R, G,\mathcal{T})$ dark energy model using the Pantheon Plus dataset. The Pantheon Plus dataset comprises 1701 SNe Ia data points in the redshift range $0.01 < z < 2.3$. The fitting process involves minimizing the $\chi^{2}$ function, which measures the deviation of the theoretical model from the observed data. The theoretical model involves the distance modulus function, which is related to the luminosity distance $d_L(z)$ as $\mu(z)=5\log_{10}\left[\frac{d_L(z)}{Mpc}\right]+25$, where $Mpc$ is the unit of distance. The luminosity distance $d_L(z)$ is defined as $d_L(z)=(1+z)\int_{0}^{z}\frac{c}{H(z)}dz'$, where $c$ is the speed of light and $H(z)$ is the Hubble parameter.\\\\

\begin{figure}[!htb]
\centering
\includegraphics[scale=0.4]{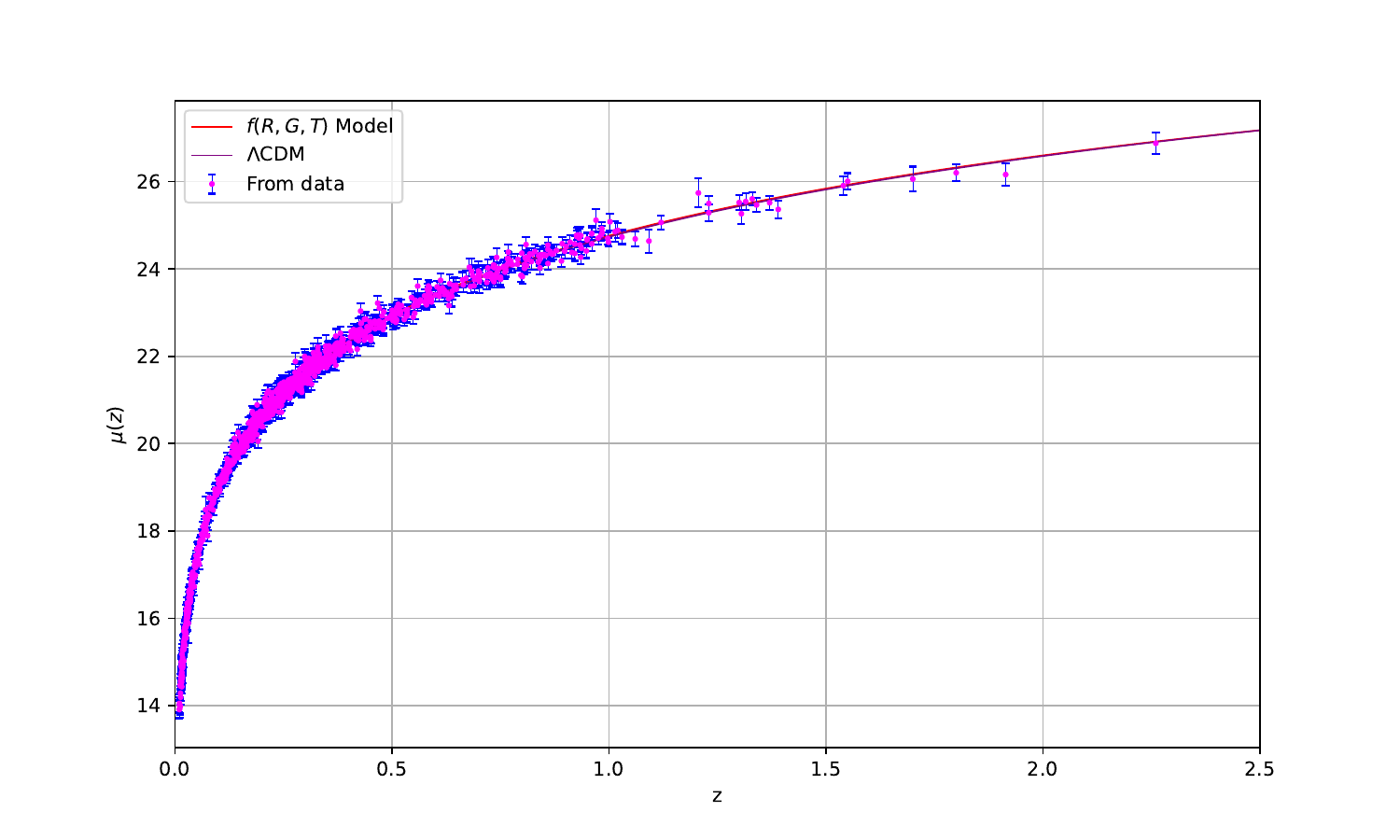}
\caption{The figure shows the theoretical curve of the distance modulus $ \mu(z) $ corresponding to the studied (red curve)  and $ \Lambda$CDM (black curve) models against the Supernovae type Ia dataset shown magenta dots with their corresponding error bars in the magenta dots with their corresponding error bars shown in blue line. For $\Lambda$CDM the parameters were fixed at $\Omega_{\mathrm{m0}} = $ 0.3 and $\Omega_\Lambda = $ 0.7.}
\label{mu(z)}
\end{figure}

Our results demonstrate that the $\Lambda$CDM model provides an excellent fit to the Pantheon Plus dataset, and $f(R, G,\mathcal{T})$ dark energy model. The comparison findings are shown in Figure \ref{mu(z)}.

\newpage
\section{Cosmographic analysis}\label{sec4}
Cosmography parameters are a set of cosmological parameters that describe the expansion history of the universe in a model-independent way. These parameters are derived solely from the measurement of the cosmic expansion rate, also known as the Hubble parameter, and its derivatives with respect to redshift. The most commonly used cosmography parameters are the Hubble constant ($H_0$), the deceleration parameter ($q_0$), and the jerk parameter ($j_0$) \cite{Visser:2004bf,Shafieloo:2012mv}. $H_0$ represents the current expansion rate of the universe, while $q_0$ and $j_0$ describe the acceleration and jerk of the expansion, respectively. Cosmography parameters have several advantages over other cosmological parameters, such as those derived from cosmic microwave background radiation or the large-scale structure of the universe. They are model-independent, meaning that they do not rely on any assumptions about the underlying cosmological model. This makes them useful for testing the validity of different cosmological models and for constraining the properties of dark energy, which is responsible for the acceleration of the expansion \cite{Lazkoz:2005sp}. Cosmography parameters are relatively easy to measure from observational data. The Hubble constant can be determined from observations of Type Ia supernovae, gravitational lensing, and other methods, while the higher-order cosmography parameters can be estimated from measurements of the Hubble parameter at different redshifts \cite{Visser:2004bf,Shafieloo:2012mv}. Cosmography parameters can be used to test the validity of the cosmological principle, which states that the universe is homogeneous and isotropic on large scales. Deviations from the cosmography predictions could indicate the presence of large-scale structures or other departures from the standard cosmological model \cite{Gomez-Valent:2018hwc}. Cosmography parameters provide a powerful tool for measuring the expansion history of the universe and testing cosmological models in a model-independent way. The ease of measurement and their ability to test the cosmological principle make them an important component of modern cosmology research.

\subsection{The deceleration parameter}
The deceleration parameter is a crucial cosmological parameter that characterizes the expansion rate of the universe. It is defined as the ratio of the deceleration of the universe's expansion to the present expansion rate. A positive deceleration parameter implies that the expansion of the universe is slowing down, while a negative deceleration parameter indicates that the expansion of the universe is accelerating. In other words, the deceleration parameter is a measure of the transition between the decelerated and accelerated phases of the universe's expansion.  Mathematically, one can define it as
\begin{equation}
q=-\frac{a\ddot{a}}{\dot{a}^2}.
\end{equation}
Observational analysis plays a critical role in determining the range of values for the deceleration parameter. For instance, the apparent brightness and redshift for supernovae in distant galaxies can be used to estimate the deceleration parameter \cite{Visser:2004bf,Shafieloo:2012mv,him1,him2,him3,him4,him5,him6,him7,him8,him9}. Recent observations strongly support models that predict an accelerating universe. However, obtaining an accurate value for the deceleration parameter remains a challenging task. It is essential to note that the Hubble parameter's behavior is determined by the sign of the deceleration parameter. If the deceleration parameter is positive, the Hubble parameter decreases with time, and if it is negative, the Hubble parameter increases with time. Therefore, the deceleration parameter provides valuable insights into the dynamics of the universe's expansion. It is crucial to explore the ranges of possible values for the deceleration parameter through observational analyses and to obtain accurate estimates to gain a better understanding of the universe's evolution.\\\\\

\begin{figure}[htbp]
\centering
\includegraphics[scale=0.42]{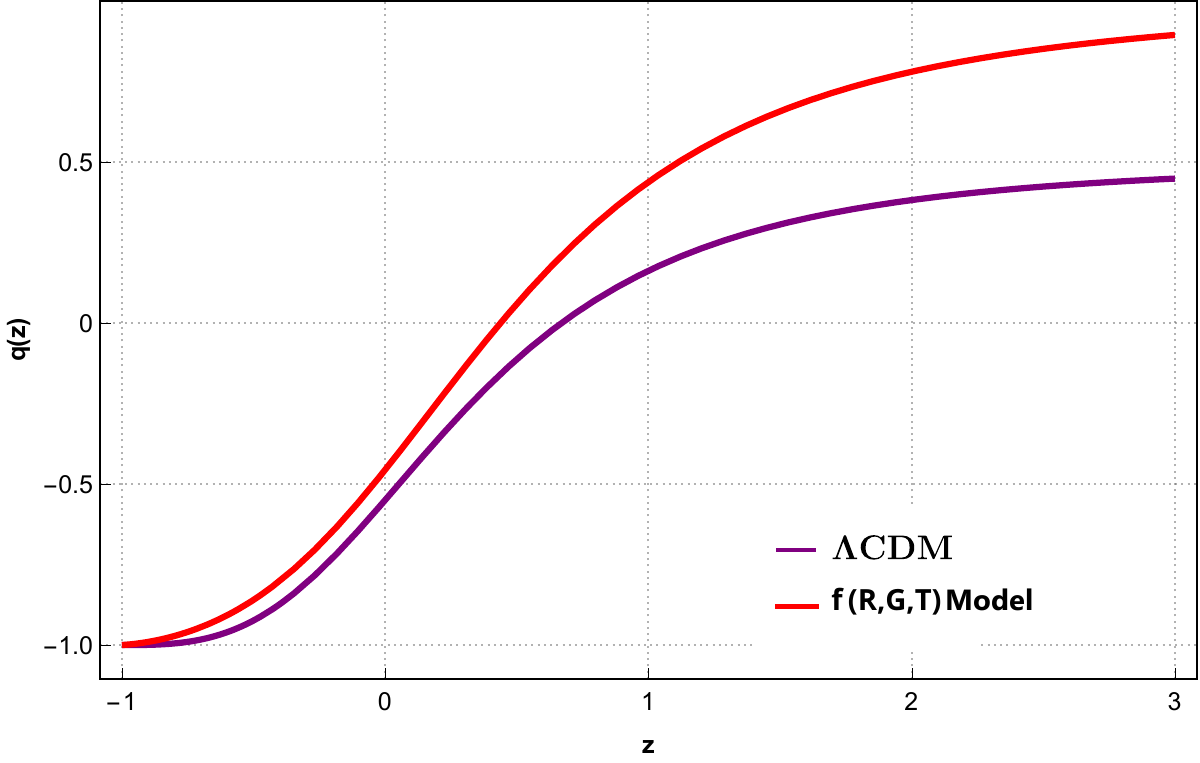}
\caption{Evolution of the deceleration parameter as a function of the redshift $z$}\label{q_z}
\end{figure}
\subsection{The jerk parameter}
The jerk parameter is a cosmological parameter that generalizes the expansion rate of the universe beyond the usual parameters of $a(t)$ and $q$. It arises from the fourth term in a Taylor series of the scale factor around a given time $t_0$ \cite{Sahni:2002fz}:

\begin{equation}
\begin{aligned}
\frac{a(t)}{a_0} & = 1+H_0(t-t_0)-\frac{1}{2}q_0H^2_{0}(t-t_0)^2+
\frac{1}{6}j_0H^3_{0}(t-t_0)^3 \\ & +O\left[(t-t_0)^4\right].
\end{aligned}
\end{equation}

The jerk parameter, denoted as $j$, is defined as the third derivative of the scale factor with respect to cosmic time \cite{Visser:2004bf}:

\begin{equation}
j=\frac{1}{a}\frac{d^3a}{d\tau
^3}\left[\frac{1}{a}\frac{da}{d\tau}\right]^{-3}=q(2q+1)+(1+z)\frac{dq}{dz}.
\end{equation}

The jerk parameter plays a significant role in the search for a suitable candidate for the physical interpretation of cosmic dynamics in the presence of various dark energy proposals \cite{Visser:2004bf}. The value of the jerk parameter can provide insight into the transitions between different eras of accelerated expansion. A specific value of the jerk parameter can establish a correspondence between dark energy proposals and standard universe models, facilitating the search for a favorable candidate for cosmic dynamics. For example, in the flat $\Lambda$CDM model, the jerk parameter has a value of $j=1$ \cite{Visser:2004bf}. The jerk parameter is a useful tool in understanding the dynamics of the universe and in distinguishing between different dark energy proposals. Its value can help us better comprehend the various eras of accelerated expansion and identify a suitable model for cosmic dynamics.
\begin{figure}[htbp]
\centering
\includegraphics[scale=0.45]{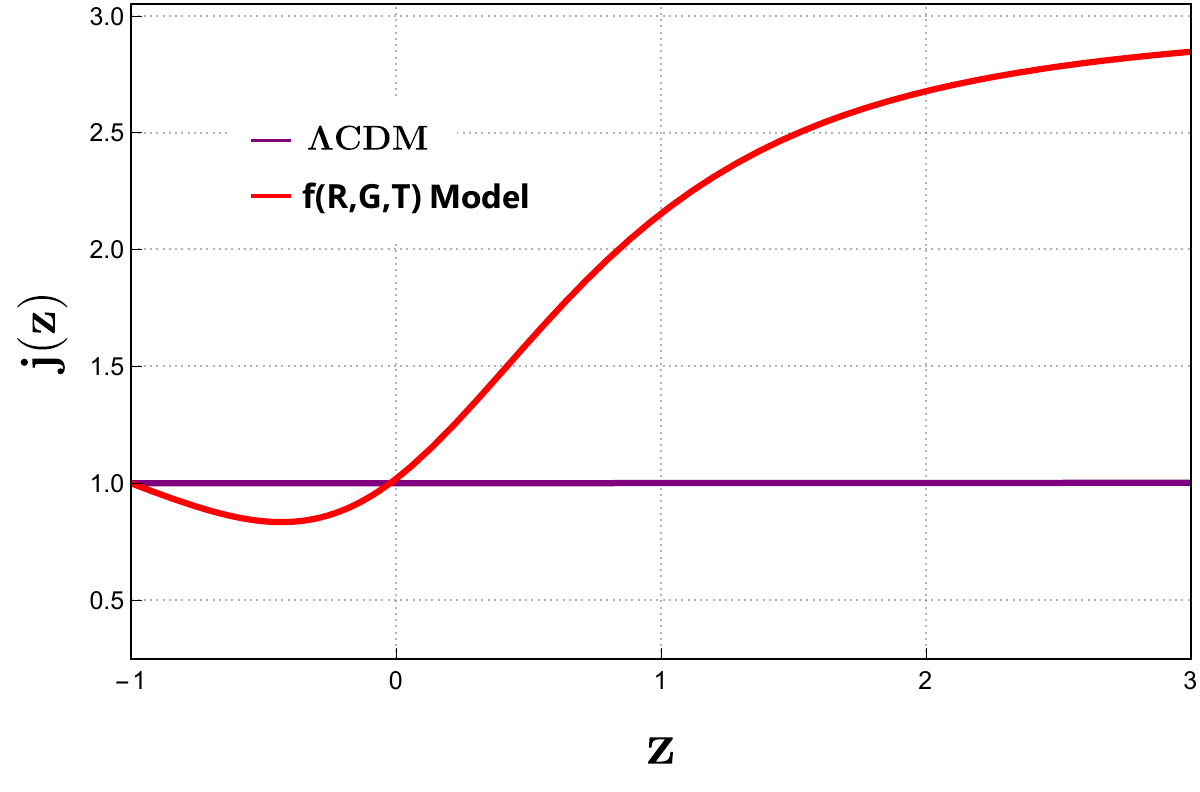}
\caption{Evolution of the jerk parameter as a function of the redshift $z$}
\label{j_z}
\end{figure}
\subsection{The snap parameter}
The Snap parameter, or the jounce parameter, is a higher-order time derivative of the expansion factor of the universe. It is the fifth term in the Taylor series expansion of the scale factor around the present time $a_0$,
\begin{equation}
\begin{aligned}
\frac{a(t)}{a_0} & =1+H_0(t-t_0)-\frac{1}{2}q_0H^2_{0}(t-t_0)^2 \\ &+\frac{1}{6}j_0H^3_{0}(t-t_0)^3+
\frac{1}{24}s_0H^4_{0}(t-t_0)^4+O\left[(t-t_0)^5\right].
\end{aligned}
\end{equation}
and it characterizes the deviation of the universe's expansion from the $\Lambda$CDM model. The snap parameter is defined as the fourth derivative of the scale factor with respect to cosmic time,
\begin{equation}
s=\frac{1}{a}\frac{d^4a}{d\tau
^4}\left[\frac{1}{a}\frac{da}{d\tau}\right]^{-4}=\frac{j-1}{3\left(q-\frac{1}{2}\right)},
\end{equation}
normalized by a certain combination of the scale factor and its time derivatives. The snap parameter plays a crucial role in characterizing the dynamics of the universe \cite{Capozziello:2003tk}. Specifically, it helps to identify the degree of deviation from the standard $\Lambda$CDM model, which assumes a cosmological constant as the source of dark energy. The snap parameter is related to the cosmic jerk parameter, and their relative behavior provides insights into the transitions between different eras of the universe's accelerated expansion. In particular, the divergence of the snap parameter with respect to the deceleration parameter determines how the universe's evolution deviates from the $\Lambda$CDM dynamics. Therefore, the snap parameter is a valuable tool for studying the nature of dark energy and its role in the evolution of the universe.\\\\\
\begin{figure}[htbp]
\centering
\includegraphics[scale=0.4]{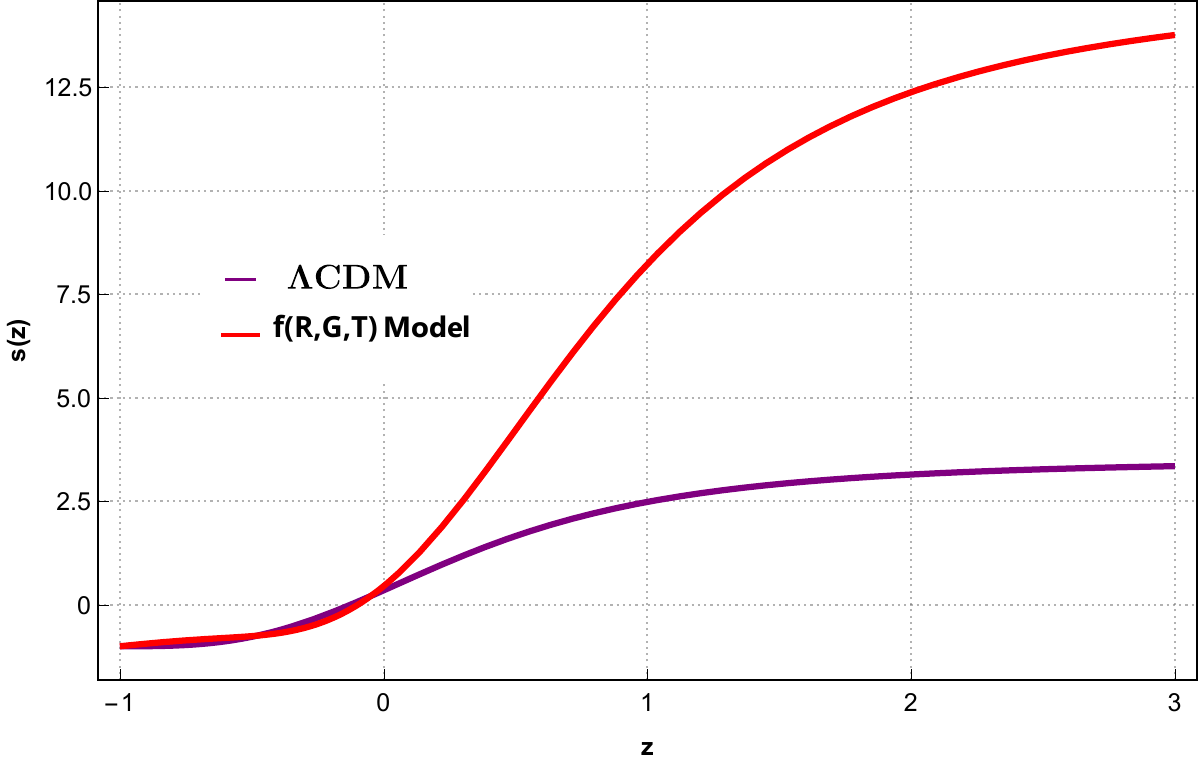}
\caption{Evolution of the snap parameter as a function of the redshift $z$}
\label{s_z}
\end{figure}
\section{Statefinder Diagnostic}\label{sec5}
The Statefinder diagnostics is a powerful tool used to study different models of dark energy (DE) and understand their characteristics based on higher-order derivatives of the scale factor. It provides a dimensionless pair of parameters, $\{r, s\}$, which can be used to analyze the cosmic properties of DE independent of specific models. The calculations for $r$ and $s$ involve expressions that involve the third derivative of the scale factor ($\dddot{a}$), the Hubble parameter ($H$), and the deceleration parameter ($q$) \cite{Sahni:2002fz}. The parameter $s$ is a linear combination of $r$ and $q$, enabling further insights into the DE behavior:

\begin{equation}
r=\frac{\dddot{a}}{a H^3}, \quad s=\frac{r-1}{3\left(q-\frac{1}{2}\right)}.
\end{equation}
Certain pairs of $r$ and $s$ have been associated with standard models of DE. For example, $\{r, s\}=\{1, 0\}$ corresponds to the $\Lambda$CDM model, while $\{r, s\}=\{1, 1\}$ corresponds to the standard cold dark matter model (SCDM) in the Friedmann-Lematre-Robertson-Walker (FLRW) universe. The range $(-\infty, \infty)$ represents the Einstein static universe. By examining the $r-s$ plane, it is possible to identify quintessence-like and phantom-like models of DE, characterized by positive and negative values of $s$, respectively. Deviations from the standard range $\{r, s\}=\{1, 0\}$ can indicate an evolutionary process from phantom-like to quintessence-like behavior \cite{Visser:2004bf}.The Statefinder diagnostics provide a comprehensive framework to explore and analyze various DE models. By employing the $\{r, s\}$ parameter pair, independent of specific DE models, one can gain insights into the cosmic behavior of DE. This diagnostic tool aids in understanding the transition between different DE phases, distinguishing between quintessence-like and phantom-like behavior, and identifying the standard DE models within the $r-s$ plane.

\begin{figure}[!htb]
   \begin{minipage}{0.49\textwidth}
     \centering
   \includegraphics[scale=0.42]{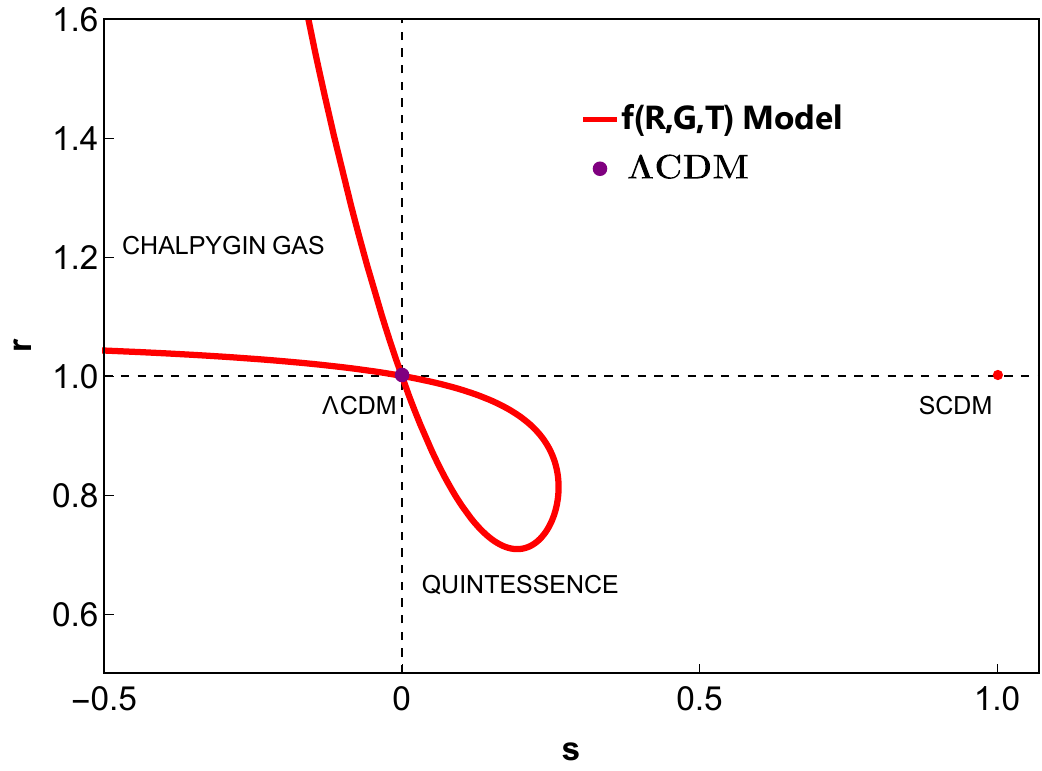}
\caption{Behavior of $\{s, r\}$ profile.}\label{sr}
   \end{minipage}\hfill
   \begin{minipage}{0.49\textwidth}
     \centering
    \includegraphics[scale=0.42]{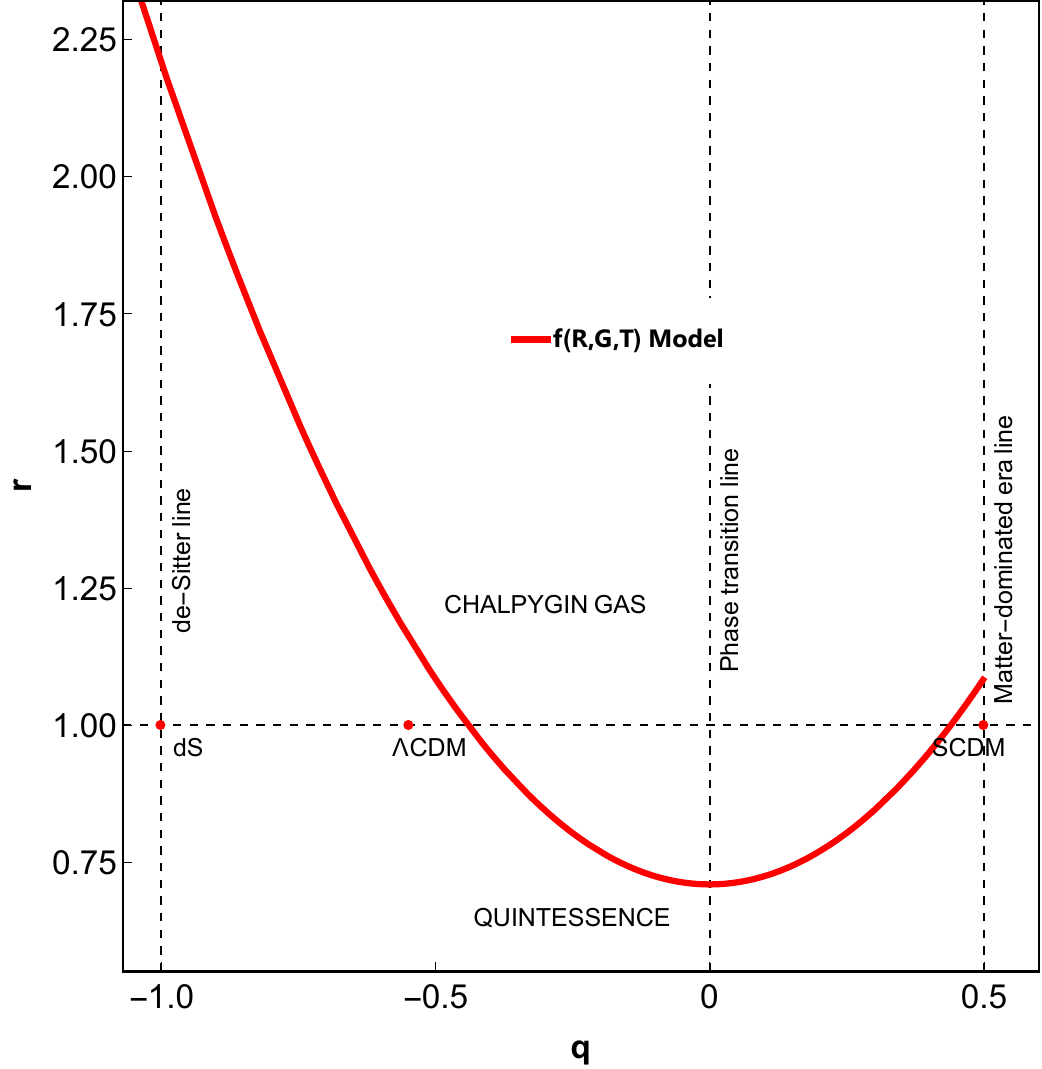}
\caption{Behavior of $\{q, r\}$ profile.}\label{qr}
   \end{minipage}
\end{figure}

\section{$Om$ Diagnostic}\label{sec6}
In cosmology, a geometrical formalism is employed in which the Hubble parameter serves as a null test for the $\Lambda$CDM model \cite{Sahni:2006pa}. Additionally, the Om diagnostic parameter is used to effectively differentiate various dark energy (DE) models from the $\Lambda$CDM model by observing the slope variation of $Om(z)$ \cite{Sahni:2008xx}. A quintessence or phantom model can be identified through a positive or negative slope of the diagnostic parameter, respectively. Furthermore, a constant slope with respect to redshift depicts a DE model corresponding to the cosmological constant. For a flat universe, the diagnostic parameter $Om(z)$ is defined as:
\begin{equation}
O m(z)=\frac{\left(\frac{H(z)}{H_0}\right)^2-1}{(1+z)^3-1}.
\end{equation}
This diagnostic parameter involves only the first-order temporal derivative, as compared to the statefinder diagnosis \cite{Sahni:2002}, which is discussed in \cite{Sahni:2008xx}. Additionally, it is applicable to Galileon models \cite{Chow:2009fm,Appleby:2009uf}, as described in \cite{Sahni:2014ooa}. Overall, the Om diagnostic parameter is a useful tool for studying the properties of dark energy models and distinguishing them from the cosmological constant model.

\begin{figure}[H]
\centering
\includegraphics[scale=0.42]{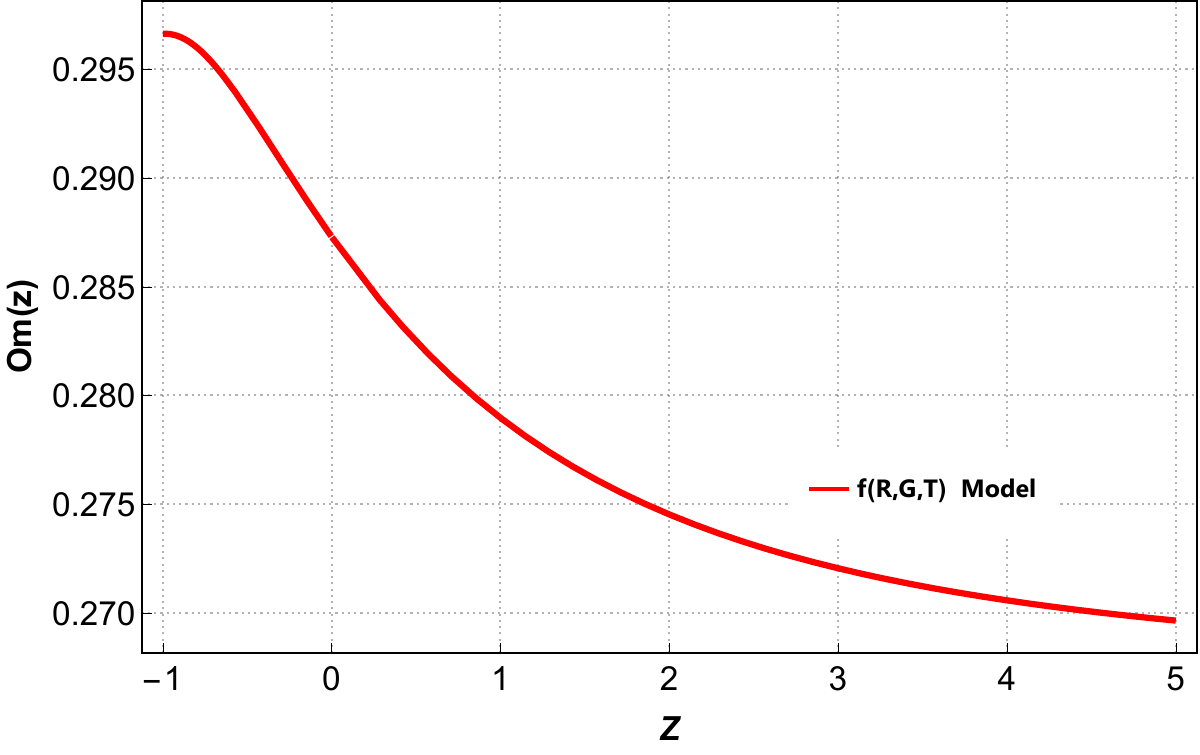}
\caption{Evolution of $Om(z)$ profile}
\label{Om}
\end{figure}


\section{Information Criteria}\label{sec7}
In cosmology, the $\chi^2_{\text{min}}$, $\chi^2_{\text{red}}$, Akaike Information Criterion (AIC), and $\Delta$AIC are commonly used statistical measures to assess the goodness of fit and compare different cosmological models based on observational data \cite{AIC1,AIC2,AIC3,AIC4}. The $\chi^2_{\text{min}}$ is defined as the minimum value of the $\chi^2$ statistic, which quantifies the difference between the observed data and the theoretical predictions of a model. It is calculated as the sum of the squared differences between the observed data points and the corresponding model predictions, divided by the measurement uncertainties:

\begin{equation}
\chi^2_{\text{min}} = \sum \frac{(O_i - E_i)^2}{\sigma_i^2},
\end{equation}
where $O_i$ and $E_i$ represent the observed and expected values, respectively, and $\sigma_i$ is the corresponding measurement uncertainty. The $\chi^2_{\text{red}}$ is the reduced chi-square statistic, obtained by dividing the $\chi^2_{\text{min}}$ by the number of degrees of freedom (NDF). The NDF is equal to the total number of data points minus the number of free parameters in the model:

\begin{equation}
\chi^2_{\text{red}} = \frac{\chi^2_{\text{min}}}{\text{NDF}}.
\end{equation}
The $\chi^2_{\text{red}}$ provides a normalized measure of the goodness of fit per degree of freedom, allowing for comparisons between models with different numbers of free parameters. The AIC is a statistical criterion that takes into account both the goodness of fit and the complexity of a model. It is calculated as:

\begin{equation}
\text{AIC} = 2k - 2\ln(\mathcal{L}),
\end{equation}
where $k$ is the number of free parameters in the model and $\mathcal{L}$ is the maximum likelihood of the model. The AIC penalizes models with more parameters, favoring simpler models that provide a good fit to the data. The $\Delta$AIC is the difference in AIC values between the two models. It is calculated as:

\begin{equation}
\Delta\text{AIC} = \text{AIC}_i - \text{AIC}_\text{min},
\end{equation}
where $\text{AIC}_i$ and $\text{AIC}_\text{min}$ are the AIC values of the $i$th model and the model with the minimum AIC value, respectively. The $\Delta$AIC provides a measure of the relative support for different models, with lower values indicating better model fit and higher likelihood . In addition to the $\chi^2_{\text{min}}$, $\chi^2_{\text{red}}$, Akaike Information Criterion (AIC), and $\Delta$AIC, another important statistical measure used in cosmology for model comparison is the Bayesian Information Criterion (BIC) \cite{BIC1,BIC2,BIC3}, along with its difference, $\Delta$BIC. The Bayesian Information Criterion (BIC) is a statistical criterion that, like AIC, takes into account both the goodness of fit and the complexity of a model. It is derived from a Bayesian perspective and is defined as:
\begin{equation}
\text{BIC} = k \ln(n) - 2\ln(\mathcal{L}),
\end{equation}
where $k$ is the number of free parameters in the model, $n$ is the number of data points, and $\mathcal{L}$ is the maximum likelihood of the model. Similar to the AIC, the BIC penalizes models with more parameters to favor simpler models that provide a good fit to the data. The difference in BIC values between two models, denoted as $\Delta$BIC, is calculated as:

\begin{equation}
\Delta\text{BIC} = \text{BIC}_i - \text{BIC}_\text{min},
\end{equation}
where $\text{BIC}_i$ and $\text{BIC}_\text{min}$ are the BIC values of the $i$th model and the model with the minimum BIC value, respectively. Like the $\Delta$AIC, the $\Delta$BIC provides a measure of the relative support for different models, and lower values of $\Delta$BIC indicate better model fit and higher likelihood. The BIC and $\Delta$BIC are additional tools that researchers use in cosmology to assess the goodness of fit and compare different models based on observational data. These criteria offer a way to strike a balance between model complexity and goodness of fit, helping researchers make informed decisions when selecting the most appropriate cosmological models. The $\chi^2_{\text{min}}$, $\chi^2_{\text{red}}$, AIC, $\Delta$AIC, BIC, and $\Delta$BIC are all essential statistical measures in cosmology for evaluating the fit of different models to observational data and for comparing models with varying degrees of complexity. These measures play a crucial role in model selection and help researchers understand which models best describe the behavior of the universe based on available data.\\\\

\begin{table}[H]
\begin{center}
\begin{tabular}{|c|c|c|c|c|c|c|}
\hline
Model & ${\chi_{\text{tot}}^2}^{min} $ & $\chi_{\text {red }}^2$ & $AIC$ & $\Delta AIC$  &  $BIC$  & $\Delta BIC$ \\
\hline $\Lambda$CDM & 1778.74 &0.934 & 1782.74 & 0 & 1799.76 & 0 \\
\hline $f(R, G, \mathcal{T})$ & 1765.22  & 0.928 & 1783.22 & -1.52 & 1828.28 & 28.5202 \\
\hline
\end{tabular}
\caption{Summary of ${\chi_{\text{tot}}^2}^{min} $, $\chi_{\text {red }}^2$, $A I C$, $\Delta A I C_c$, $B I C$, $\Delta B I C$ for  $\Lambda$CDM and $f(R, G, \mathcal{T})$ model} \label{table2}
\end{center}
\end{table}

\section{Results}\label{sec8}
\paragraph{Deceleration parameter:}
The fig \ref{q_z} represents the behavior of the deceleration parameter ($q$) for the $f(R, G, \mathcal{T})$ model and the $\Lambda$CDM model at different epochs and the phase transition redshift. At high redshifts ($z\to\infty$), the $f(R, G, \mathcal{T})$ model has a deceleration parameter value of $0.824$, suggesting a decelerating expansion of the universe. On the other hand, the $\Lambda$CDM model exhibits a slightly lower deceleration parameter value of $0.451$, indicating a comparatively slower deceleration. As we move towards the present epoch ($z\to 0$), both models show a transition towards a negative deceleration parameter. The $f(R, G, \mathcal{T})$ model has a deceleration parameter value of $-0.494$, indicating a transition to an accelerating phase. Similarly, the $\Lambda$CDM model exhibits a slightly lower value of $-0.555$, also signifying a transition towards acceleration. At the cosmological constant limit ($z\to -1$), both models converge to a deceleration parameter value of $-1$, indicating a transition to a de Sitter phase where the expansion is accelerating at an ever-increasing rate. Lastly, the figure provides the phase transition redshift ($z_{tr}$) at which the deceleration parameter for each model reaches zero ($q=0$). For the $f(R, G, \mathcal{T})$ model, the transition occurs at $z_{tr}=0.4253$, while for the $\Lambda$CDM model, the transition occurs at a slightly higher redshift of $z_{tr}=0.659$. Overall, the figure demonstrates the different behaviors of the deceleration parameter between the $f(R, G, \mathcal{T})$ model and the $\Lambda$CDM model at various epochs and the phase transition redshift. It illustrates how the expansion of the universe transitions from a decelerating phase to an accelerating phase, with the $f(R, G, \mathcal{T})$ model showing slightly different values compared to the $\Lambda$CDM model at each epoch.\\\\\

\paragraph{Jerk parameter:}
The fig \ref{j_z} presents the behavior of the jerk parameter ($j$) for the $f(R, G,\mathcal{T})$ model and the $\Lambda$CDM model at different epochs and the cosmological constant limit ($z\to -1$). At high redshifts ($z\to\infty$), The $f(R, G, \mathcal{T})$ model exhibits a value of $2.7345$ for the jerk parameter ($j$). On the other hand, the $\Lambda$CDM model maintains a jerk parameter value of 1 at this epoch. This distinction suggests that at very high redshifts, the $f(R, G, \mathcal{T})$ model differs from the $\Lambda$CDM model in terms of the rate of change of acceleration. The $f(R, G, \mathcal{T})$ model exhibits a higher value for the jerk parameter, indicating a potentially different dynamic behavior compared to the constant rate of change of acceleration observed in the $\Lambda$CDM model. As we approach the present epoch ($z\to 0$), the $f(R, G,\mathcal{T})$ model exhibits a jerk parameter value of 1, suggesting a constant acceleration similar to the $\Lambda$CDM model.
At the cosmological constant limit ($z\to -1$), both models have a jerk parameter value of $1$. This implies that the acceleration remains constant in both models at this limit.\\\\\

\paragraph{Snap parameter:}
The  fig \ref{s_z} presents the behavior of the snap parameter ($s$) for the $f(R, G,\mathcal{T})$ model and the $\Lambda$CDM model at different epochs and the cosmological constant limit ($z\to -1$). At high redshifts ($z\to\infty$), the snap parameter for the $f(R, G,\mathcal{T})$ Model is $13.353$, while for the $\Lambda$CDM model, it is $2.634$. This indicates a significant deviation in the curvature and evolution of the two models during the early stages of the universe. The $f(R, G,\mathcal{T})$ Model exhibits a higher snap parameter, suggesting a more pronounced curvature and evolution compared to the $\Lambda$CDM model. As we approach the present epoch ($z\to 0$), the snap parameter for both models decreases. For the $f(R, G,\mathcal{T})$ model, it becomes $0.1685$, while for the $\Lambda$CDM model, it is $0.1674$. This implies that both models converge to a similar behavior in terms of curvature and evolution at the current epoch. At the cosmological constant limit ($z\to -1$), both models have a snap parameter of $-0.634$. This indicates a transition to an accelerated expansion phase, where the universe's curvature and evolution are primarily governed by the cosmological constant. The comparative analysis of the snap parameter for the $f(R, G,\mathcal{T})$ model and the $\Lambda$CDM model reveals that the two models exhibit different behaviors in terms of curvature and evolution at high redshifts, with the $f(R, G,\mathcal{T})$ model showing a more pronounced curvature. However, as we approach the present epoch, both models converge to a similar behavior. At the cosmological constant limit, both models exhibit an accelerated expansion phase.\\\\\

\paragraph{The $\{s, r\}$ Profile:}
The fig ~\ref{sr} provides a comprehensive description of $\{s,r\}$ profile of the Statefinder diagnostic. At early times, the $f(R, G,\mathcal{T})$ model is found to have values of $r > 1$ and $s < 0$. These values indicate that the model resembles a Chaplygin gas-type dark energy model. The Chaplygin gas is a theoretical construct that can describe the behavior of dark energy, characterized by an equation of state that deviates from the standard cosmological constant. As the universe evolves, the $f(R, G,\mathcal{T})$ model transitions to values between $r < 1$ and $s > 0$, indicating a quintessence domain. Quintessence refers to a form of dark energy that possesses a dynamic nature, often associated with a slowly evolving scalar field. The quintessence domain represents a phase in which the dark energy density remains relatively constant over time. During this evolution, the $f(R, G,\mathcal{T})$ model crosses an intermediate fixed point, represented by the $\Lambda$CDM point $\{1,0\}$. The $\Lambda$CDM model is the standard cosmological model that includes a cosmological constant ($\Lambda$) and cold dark matter (CDM). Crossing this fixed point suggests a transition from one phase to another, from a Chaplygin gas-like behavior to a quintessence-like behavior. Finally, at late times, the $f(R, G,\mathcal{T})$ model returns to the Chaplygin gas domain by passing an interim fixed $\Lambda$CDM point. This indicates a transition back to a Chaplygin gas-type dark energy behavior. The described model exhibits a rich cosmological evolution, transitioning between different phases of dark energy behavior. It begins as a Chaplygin gas-type model, evolves through a quintessence domain while crossing an intermediate $\Lambda$CDM fixed point, and eventually returns to the Chaplygin gas-type behavior at late times.\\\\

\paragraph{The $\{q, r\}$ Profile:}
The fig ~\ref{qr} provides a comprehensive description of $\{q,r\}$ profile of the Statefinder diagnostic. At early times, the model exhibits values in the range $q < 0$ and $r < 1$ corresponds to the quintessence domain. As the model evolves, it transitions to values where $r > 1$ and $q < 0$. This region indicates the Chaplygin gas domain. The transition from the quintessence domain to the Chaplygin gas domain signifies a change in the dominant energy component of the universe, potentially leading to different cosmological behaviors. Finally, at late times, the $f(R, G,\mathcal{T})$ model deviate towards the de Sitter line, represented by q $=$-1. The de Sitter line corresponds to a universe dominated by dark energy, resulting in accelerated expansion and a constant equation of state parameters. Deviation towards this line indicates a pure de Sitter universe, potentially due to the influence of additional cosmological components or modified gravity. Overall, the $\{q,r\}$ profile of the $f(R, G,\mathcal{T})$ energy model reveal a rich cosmological evolution. The model transitions from a quintessence domain to a Chaplygin gas domain, indicating changes in the dominant energy component. Departure from the de Sitter line at late times suggests the presence of additional effects or modifications to the standard cosmological model.\\\\

\paragraph{$Om$ Diagnostic:}
The $Om$ Diagnostic is a cosmological quantity that helps characterize the behavior of the matter density in the universe. The Fig ~\ref{Om} provided presents the values of the $Om$ Diagnostic for  $f(R, G,\mathcal{T})$ model at different epochs represented by the redshift $z$. At $z\to\infty$, which corresponds to the early universe or the farthest observable regions, the $Om$ Diagnostic for  $f(R, G,\mathcal{T})$ is recorded as $0.270$. This value indicates the matter density contribution to the overall energy density of the universe at very high redshifts. As we approach the present time, represented by $z\to 0$, the $Om$ Diagnostic for  $f(R, G,\mathcal{T})$ model increases to $0.2855$. This suggests that matter density plays a slightly more significant role in the total energy density of the universe as compared to the early universe. Finally, at the phase transition redshift $z\to -1$, which signifies the late-time universe, the $Om$ Diagnostic reaches a value of $0.2952$. This indicates that matter density continues to have a non-negligible contribution to the total energy density of the universe, even in the late stages of cosmic evolution. Overall, the values of the $Om$ Diagnostic for  $f(R, G,\mathcal{T})$ model suggest that matter density remains an important component of the energy content of the universe throughout its history, from the early universe to the present and even in the late-time universe.\\\\

\paragraph{Information Criteria:} Based on the values presented in Table \ref{table2}, we can provide a comprehensive comparison between the $f(R, G, \mathcal{T})$ and $\Lambda$CDM models using various statistical measures, including ${\chi_{\text{tot}}^2}^{min}$, $\chi_{\text{red}}^2$, $AIC$, $\Delta AIC_c$, $BIC$, and $\Delta BIC$. The ${\chi_{\text{tot}}^2}^{min}$ for the $f(R, G, \mathcal{T})$ model is slightly lower than that of the $\Lambda$CDM model, indicating a slightly better overall fit of the $f(R, G, \mathcal{T})$ model to the data. The reduced $\chi_{\text{red}}^2$ values for both models are very close, suggesting that both models provide reasonable fits to the data. The $AIC$ value for the $f(R, G, \mathcal{T})$ model is higher than that of the $\Lambda$CDM model, indicating that the $\Lambda$CDM model has a better balance between goodness of fit and model complexity according to AIC. The calculated $\Delta AIC_c$ of $-1.52$ suggests that the $f(R, G, \mathcal{T})$ model is strongly favored over the $\Lambda$CDM model. The $BIC$ value for the $f(R, G, \mathcal{T})$ model is higher than that of the $\Lambda$CDM model, reinforcing the idea that the $\Lambda$CDM model is preferred in terms of model complexity and goodness of fit according to BIC. The $\Delta BIC$ value of $28.5202$ strongly favors the $\Lambda$CDM model. This indicates a substantial preference for the $\Lambda$CDM model over the $f(R, G, \mathcal{T})$ model based on BIC and $\Delta BIC$. The evidence strongly supports the $\Lambda$CDM model over the $f(R, G, \mathcal{T})$ model. The $BICc$ and $\Delta BIC$ values consistently indicate that the $\Lambda$CDM model is strongly favored in terms of goodness of fit and model selection. While the $f(R, G, \mathcal{T})$ model shows a slightly better fit in terms of ${\chi_{\text{tot}}^2}^{min}$, the overall considerations from AIC, BIC and $\Delta BIC$ strongly point to the superiority of the $\Lambda$CDM model.

\section{Discussions and Concluding Remarks}\label{sec10}
In this study, we have investigated the behavior and evolution of the $f(R, G, \mathcal{T})$ model. By analyzing various cosmological parameters and diagnostics, we have gained insights into the characteristics and dynamics of this model compared to the well-established $\Lambda$CDM model. Our analysis of the deceleration parameter ($q$) reveals that the $f(R, G, \mathcal{T})$ model exhibits a transition from a decelerating phase to an accelerating phase. The analysis of jerk parameter ($j$) demonstrates that the $f(R, G, \mathcal{T})$ model exhibits a higher rate of change of acceleration at high redshifts compared to the $\Lambda$CDM model. However, as we approach the present epoch and the cosmological constant limit, both models converge to a constant rate of change of acceleration. The analysis of snap parameter ($s$)  reveals that the $f(R, G, \mathcal{T})$ model exhibits a more pronounced curvature and evolution at high redshifts compared to the $\Lambda$CDM model. However, as we approach the present epoch, both models converge to a similar behavior, indicating a transition to an accelerated expansion phase. The analysis of Statefinder parameters provides valuable insights into the nature of dark energy and its transition between different phases in the $f(R, G, \mathcal{T})$ model. The model exhibits transitions from a Chaplygin gas-like behavior to a quintessence-like behavior, crossing an intermediate $\Lambda$CDM fixed point. This rich cosmological evolution highlights the dynamic nature of dark energy in the $f(R, G, \mathcal{T})$ model. The analysis of $Om$ diagnostic indicates that matter density remains an important component of the energy content of the universe throughout its history, including the early universe, the present epoch, and even the late-time universe. Furthermore, the comparison of information criteria suggests that the $\Lambda$CDM model performs slightly better than the $f(R, G, \mathcal{T})$ model in terms of goodness of fit and model selection.\\\\
In conclusion, the $f(R, G, \mathcal{T})$ model exhibits interesting and distinct features compared to the well-established $\Lambda$CDM model. Its dynamics, including the transition from a decelerating phase to an accelerating phase. While further investigations and analyses are necessary to validate the model's consistency with other observational data and theoretical expectations, the  obtained results in this study suggest that the $f(R, G, \mathcal{T})$ model has the potential to revolutionize the field of cosmology. Its ability to capture and explain various cosmological phenomena opens up new avenues for exploring the fundamental principles governing the evolution and behavior of our universe. The $f(R, G, \mathcal{T})$ model, with its unique characteristics and dynamics, offers a promising framework for advancing our understanding of the universe. Future studies could focus on further constraining the model parameters using a broader range of observational data, such as measurements of cosmic microwave background radiation, large-scale structure formation, and gravitational wave events. Additionally, it would be valuable to investigate the implications of the $f(R, G, \mathcal{T})$ model for other cosmological phenomena, such as the formation and evolution of galaxies, and the behavior of dark matter. Furthermore, exploring the theoretical foundations of the $f(R, G, \mathcal{T})$ model could provide deeper insights into the underlying physics that drives the modifications to general relativity and the interplay between matter density, curvature, and dark energy. This could involve investigating the model in the context of quantum gravity theories, studying its implications for particle physics, and exploring its connections to other modified gravity theories. Moreover, it would be valuable to investigate the $f(R, G, \mathcal{T})$ model in the context of other astrophysical and cosmological observations, such as the study of gravitational lensing, the cosmic microwave background polarization, and the behavior of cosmic voids. These investigations could provide additional constraints and tests for the model, helping to determine its viability and consistency with a wide range of observational data. In conclusion, the $f(R, G, \mathcal{T})$ model represents a promising avenue for advancing our understanding of the universe. By further investigating its behavior, constraining its parameters, and exploring its implications for various astrophysical and cosmological phenomena, we can deepen our understanding of the fundamental principles that govern our universe and potentially uncover new insights into its evolution and dynamics. The findings in our analysis
indicate that $f(R, G, \mathcal{T})$ gravity model may be served as a good candidate for gravitational modifications.\\\\

\textbf{Public Library}
The Python library used for plotting confidence contours can be found at \hyperlink{https://samreay.github.io/ChainConsumer/}{https://samreay.github.io/ChainConsumer/}\\\\

\textbf{ Acknowledgement:} NUM is thankful to CSIR, Govt. of India for providing Senior Research Fellowship (No. 08/003(0141))/2020-EMR-I). G. Mustafa is very thankful to Prof. Gao Xianlong from the Department of Physics, Zhejiang Normal University, for his kind support and help during this research. Further, G. Mustafa acknowledges Grant No. ZC304022919 to support his Postdoctoral Fellowship at Zhejiang Normal University.\\

\end{document}